\documentclass[aps,pre,twocolumn,showpacs]{revtex4-1}
\usepackage{graphicx,amssymb,latexsym,amsmath}
\begin{document}
\def\be{\begin{equation}}
\def\ee{\end{equation}}
\def\bea{\begin{eqnarray}}
\def\eea{\end{eqnarray}}
\def\fr{\frac}
\def\th{\theta}
\def\vp{\theta}
\def\l{\label}
\def\e{\epsilon}
\def\epsilon{\varepsilon}
\def\la{\langle}
\def\ra{\rangle}

\title{\mbox{Ensemble inequivalence in a mean-field XY model with
ferromagnetic and nematic couplings}}
\author{\mbox{Arkady Pikovsky$^{1,2}$, Shamik Gupta$^3$, Tarcisio N. Teles$^4$,
Fernanda P. C. Benetti$^4$, Renato Pakter$^4$, Yan Levin$^4$, Stefano
Ruffo$^5$}}
\affiliation{$^1$\mbox{Department of Physics and Astronomy, Potsdam University,
Karl-Liebknecht-Str 24, D-14476, Potsdam, Germany}\\ 
$^2$\mbox{Department of Control Theory, Nizhni Novgorod State University, 
Gagarin Av. 23, 606950, Nizhni Novgorod, Russia} \\
\mbox{$^3$Laboratoire de Physique Th\'{e}orique et Mod\`{e}les
Statistiques (CNRS UMR 8626), Universit\'{e} Paris-Sud, Orsay, France}\\
\mbox{$^4$Instituto de F\'{i}sica, Universidade Federal do Rio Grande do Sul,
Caixa Postal 15051, CEP 91501-970, Porto Alegre, RS, Brazil}
\mbox{$^5$Dipartimento di Fisica e Astronomia and CSDC, Universit\`{a} di Firenze, INFN and
CNISM, 1 50019 Sesto Fiorentino, Italy}
}
\begin{abstract}
We explore ensemble inequivalence in long-range interacting systems by
studying an XY model of classical spins with ferromagnetic and nematic
coupling. We
demonstrate the inequivalence by mapping the microcanonical phase diagram
onto the canonical one, and also by doing the inverse mapping. We show that the equilibrium phase
diagrams within the two ensembles strongly disagree
within the regions of first-order transitions, exhibiting interesting
features like temperature jumps. In particular,
we discuss the coexistence and forbidden regions of different macroscopic states in both the phase diagrams. 
\end{abstract}
\date{\today}
\maketitle

Recent years have seen extensive studies of systems with long-range
interactions that have the two-body potential in $d$ dimensions decaying at large separation $r$
as $1/r^{\alpha}; 0\le \alpha \le d$ \cite{Campa:2009,Bouchet:2010,Levin:2014,Campa:2014}. Examples
span a wide variety, from bacterial population \cite{Sopik:2005}, plasmas \cite{Nicholson:1992}, dipolar
ferroelectrics and ferromagnets \cite{Landau:1960}, to two-dimensional geophysical
vortices \cite{Chavanis:2002}, self-gravitating
systems \cite{Paddy:1990}, etc. A striking feature of long-range systems
distinct from short-range ones is that of non-additivity, whereby thermodynamic quantities scale superlinearly
with the system size. Non-additivity manifests in static properties like
negative microcanonical specific heat
\cite{Lynden-Bell:1968,Thirring:1970}, inequivalence of statistical
ensembles
\cite{Kiessling:1997,Barre:2001,Ispolatov:2001,Barre:2007,Mukamel:2005,Venaille:2009,Venaille:2011,Teles:2014},
and other rich possibilities \cite{Bouchet:2005}. As for the dynamics, long-range systems often exhibit broken
ergodicity \cite{Mukamel:2005,Bouchet:2008}, and slow relaxation towards equilibrium \cite{Chavanis:2002,Mukamel:2005,Ruffo:1995,Yamaguchi:2004,Campa:2007,Joyce:2010}.

Here, we demonstrate ensemble inequivalence in a model of long-range
systems that has mean-field
interaction (i.e., $\alpha=0$) and two coupling modes. 
This so-called Generalized Hamiltonian Mean-Field (GHMF) model, a
long-range version with added kinetic energy of the model of Ref. \cite{Lee:1985}, has $N$
interacting particles with angular coordinates $\th_i
\in [0,2\pi]$ and momenta $p_i$, $i=1,2,\ldots,N$, which are moving on
a unit circle \cite{Teles:2012}. The GHMF Hamiltonian is 
\be
H=\sum\limits_{i=1}^N \fr{p_i^2}{2}+\fr{1}{2N}\sum\limits_{i,j=1}^N\Big[1-\Delta
\cos \th_{ij}-(1-\Delta)\cos 2\th_{ij}\Big],
\l{eq:H}
\ee
where $\th_{ij}\equiv \th_i-\th_j$. Here, $\cos\th_{ij}$ is an attractive interaction minimized by the particles
forming a cluster, so that $\th_{ij}=0\; (\text{mod }2\pi)$, while
$\cos2\th_{ij}$ with two minima at $\th_{ij}=0,\,\pi\; (\text{mod }2\pi)$ promotes a two-cluster state. The parameter $\Delta \in [0,1]$ sets the relative strength of the two coupling modes. 
The potential energy in (\ref{eq:H}) is scaled by $N$ to
make the energy extensive, following the Kac prescription
\cite{Kac:1963}, but the system remains non-additive.
In terms of the $XY$-spin
vectors ${\bf S}_i\equiv(\cos~\th_i,\sin~\th_i)$, the interactions have
the form of a mean-field ferromagnetic interaction $\sim -\Delta{\bf S}_i\cdot{\bf S}_j$, 
 and a mean-field coupling $\sim -(1-\Delta)({\bf S}_i\cdot{\bf S}_j)^2$
 promoting nematic ordering. For $XY$ lattice models
with this type of ferro-nematic coupling, see~\cite{Lee:1985,Carpenter:1989,Park:2008,Qi:2013}.
The system (\ref{eq:H}) has Hamilton dynamics:
$d\th_i/dt=p_i, dp_i/dt=-\partial H/\partial \th_i$.
For $\Delta=1$, when no nematic ordering exists, the GHMF model
becomes the Hamiltonian mean-field (HMF) model \cite{Ruffo:1995}, a paradigmatic model of long-range systems
\cite{Campa:2009}.

In this work, we report on striking and strong inequivalence of statistical ensembles for the
GHMF model. The system has three equilibrium phases: ferromagnetic, paramagnetic, and nematic, with first and second-order
transitions. Let us note that ref. \cite{Antoni:2002} studied
another model with long-range interactions, which also shows
paramagnetic, ferromagnetic and nematic-like phases.
For the GHMF model, by comparing the phase diagrams in the
canonical and microcanonical ensembles (the latter is derived
in~\cite{Teles:2012}), we show in the regions of first-order
transitions that the phase diagrams differ significantly. We analyze the
inequivalence in two ways, by mapping the microcanonical phase diagram
onto the canonical one, as is usually done
\cite{Barre:2001,Ispolatov:2001,Barre:2007,Mukamel:2005,Venaille:2009,Venaille:2011,Teles:2014},
and also by doing the inverse mapping of the canonical onto the
microcanonical one; in particular, we discuss the coexistence and forbidden
regions of different macroscopic states. This study
 demonstrates the subtleties and intricacies of the presence of different stability
 regions of macroscopic states in long-range systems in
 microcanonical and canonical equilibria. It is worth noting that
 compared to the the pure para-ferro transition, the phenomenology here
 due to presence of the additional nematic phase is much more rich. We
 will show that the region where the three phases meet, within both
 microcanonical and canonical ensembles, is the one exhibiting ensemble inequivalence.

We now turn to derive our results. Rotational symmetry of the
Hamiltonian (\ref{eq:H}) allows 
to choose, without loss of generality, the ordering direction in the
equilibrium stationary state to be along
$x$ (there are no stationary states with a non-zero angle between
the directions of ferromagnetic and nematic order), and to define as order parameters the equilibrium
averages
\be
R_m \equiv \la \cos~m\th \ra; ~~m=1,2, 
\ee
where $m=1$ (respectively, $2$) stands for the ferromagnetic
(respectively, nematic) order.
The canonical
partition function is $Z=\prod_i \int dp_i d\th_i \exp(-\beta H)$, with
$\beta=1/T$ being the inverse of the temperature $T$ measured in units of the
Boltzmann constant. Since Eq.~(\ref{eq:H}) is a
mean-field system, in the thermodynamic limit $N \to \infty$, one
follows the standard Hubbard-Stratonovich
transformation and a saddle-point approximation to evaluate $Z$
\cite{Campa:2009}. One then obtains expressions for $R_m$'s, and the average energy per particle, given by
$\langle \epsilon \rangle=\partial (\beta f)/\partial \beta$, where $f$ is the free
energy per particle. One has, with $m=1,2$,
\be
R_m=\fr{\int d\th~\cos~m\th~
e^{\beta [\Delta R_1 \cos~\th+(1-\Delta) R_2 \cos~2\th]}}{\int d\th~
e^{\beta [\Delta R_1 \cos~\th+(1-\Delta) R_2 \cos~2\th]}},
\l{eq:scmfcan}
\ee
$\langle \epsilon
\rangle=1/(2\beta)+1/2-(1/2)\left(\Delta R_1
^2+(1-\Delta)R_2^2\right)$, and 
\bea
f&=&-\fr{1}{2\beta}\ln\Big(\fr{2\pi}{\beta}\Big)+\fr{1}{2}+\fr{1}{2}\left(\Delta
R_1^2+(1-\Delta)R_2^2\right)\nonumber \\
&&-\fr{1}{\beta}\ln \left(\int d\th~ e^{\beta [\Delta R_1 \cos~\th+(1-\Delta)R_2
\cos~2\th]}\right).
\l{eq:freeen}
\eea
The canonical phase diagram in the $\Delta$ -- $T$ plane is obtained by
plotting the equilibrium values of $R_1$ and $R_2$ that solve Eq.~(\ref{eq:scmfcan}) and minimize the free energy (\ref{eq:freeen}). 

We now describe a practical way to obtain the canonical phase
diagram, by
 introducing auxiliary variables
$R$, $\alpha$, as 
\bea
\begin{cases}
R \equiv \sqrt{(\beta\Delta R_1)^2+(\beta(1-\Delta) R_2)^2},\\
\cos~\alpha \equiv \beta\Delta R_1/R,\quad \sin~\alpha \equiv
\beta(1-\Delta) R_2 /R\;.
\end{cases}
\label{eq:aux}
\eea
Then, the argument of the exponential in
Eq.~(\ref{eq:scmfcan}) becomes $R(\cos\alpha~
\cos\th+\sin\alpha~
\cos2\th)$, and the integrals on the right hand side of
Eq.~(\ref{eq:scmfcan}) evaluate to two quantities $C_m(R,\alpha)$ that depend on the introduced auxiliary variables. Using $R_m=C_{m}(R,\alpha)$ we obtain,
by virtue of Eq.~(\ref{eq:aux}), all the 
parameters in a parametric form in terms of the introduced auxiliary
variables:
\bea
\beta=\fr{R\cos~\alpha}{C_1}+\fr{R\sin~\alpha}{C_2},\quad
\Delta=1-T\fr{R\sin~\alpha}{C_2}.
\label{eq:cantd}
\eea
Once $R_{1,2},\beta,\Delta$ are determined, one can use
Eq.~(\ref{eq:freeen}) to find the free energy of 
the solution. Varying $R\geq 0$ and $\alpha
\in [0,\pi/2)$ give all solutions of
Eq.~(\ref{eq:scmfcan}),
while Eq.~(\ref{eq:freeen}) yields the stable branches.
We note that in Ref.~\cite{Komarov:2013} studying a
nonequilibrium version of our model, a different and more useful method of finding $C_{1,2}$, based on the Fourier mode representation
of an equivalent Fokker-Planck equation, is used; in our equilibrium
setup, however, exploiting the integrals (\ref{eq:scmfcan}) is simpler.
For the pure nematic phase (that has $R_1=0$), 
one sets $\alpha=\pi/2$, so that
the only auxiliary parameter is $R$; one finds $R_2=C_2(R)$
from Eq.~(\ref{eq:scmfcan}), and the temperature from
$\beta=R/(R_2(1-\Delta))$.

In contrast to Eq.~(\ref{eq:scmfcan}), the order parameters within a
microcanonical ensemble, derived in~\cite{Teles:2012}, satisfy
\bea
R_{m}=\fr{\int d\th~\cos~m\th~
\exp\left[\fr{\Delta R_1 \cos~\th+(1-\Delta) R_2 \cos~2\th}{q(\e)}\right]}
{\int d\th~ \exp\left[\fr{\Delta R_1 \cos~\th+(1-\Delta) R_2
\cos~2\th}{q(\e)}\right]}.
\label{eq:scmfmc}
\eea
Here, $\e$ is the energy per particle, and $q(\e)\equiv 2\e -1+\Delta R_1^2+(1-\Delta)R_2^2$.
For given values of $\e$ and $\Delta$, the equilibrium values of $R_1$
and $R_2$ are obtained as a particular solution of Eq.~(\ref{eq:scmfmc}) that maximizes the entropy~\cite{Teles:2012}
\bea
&&s(\e)=\fr{1}{2}\ln{2\pi}+\fr{1}{2}+\fr{\ln
q(\e)}{2}-\fr{1}{2}\Big(\fr{\Delta R_1^2+(1-\Delta)R_2^2}{q(\e)}\Big)
\nonumber \\
&&+\ln \int d\th~ 
\exp\left[\fr{\Delta R_1
\cos~\th+(1-\Delta)R_2\cos~2\th}{q(\e)}\right].
\l{eq:entrop}
\eea

The averages (\ref{eq:scmfmc}) are the same as (\ref{eq:scmfcan}) on
making the identification of the microcanonical energy $\e$ with the average energy
$\langle \e \rangle$ in the canonical ensemble, so that the inverse temperature $\beta$ in
(\ref{eq:scmfcan}) is 
\begin{equation}
\beta^{-1}=q(\e)=2\e-1+\Delta R_1^2 +(1-\Delta)R_2^2.
\label{eq:beta-e}
\end{equation}
This constitutes a link between the phase diagrams in the two ensembles.
Using then the integrals (\ref{eq:scmfcan}), we get the following parametric 
representation in the $\Delta$ -- $\e$ plane for
the microcanonical ensemble: After finding $R_1=C_1(R,\alpha)$ and
$R_2=C_2(R,\alpha)$, we get $R\cos~\alpha=\Delta R_1/q(\e),R\sin~\alpha=(1-\Delta) R_2/q(\e)$,
or, explicitly, 
\bea
\begin{cases}
\Delta=\fr{R_2\cos~\alpha}{R_2\cos~\alpha+R_1\sin~\alpha},
\\
\e=\fr{1}{2}\Big[\fr{(1-\Delta) R_2}{R\sin~\alpha}-\Delta
R_1^2-(1-\Delta)R_2^2+1\Big].
\end{cases}
\label{eq:mcde}
\eea
Once $R_{1,2},\e,\Delta$ have been determined, one can use
Eq.~(\ref{eq:entrop}) to find the entropy of 
the solution. For the pure nematic phase, $\alpha=\pi/2$, 
and $R_2^2=1+(1-2\e)/(1-\Delta)$.

Summarizing, expressions (\ref{eq:scmfcan},\ref{eq:cantd}) and (\ref{eq:scmfmc},\ref{eq:mcde})
provide self-consistent stationary state solutions for the order parameters
in the canonical and the
microcanonical ensemble, respectively. 
Stable branches of these solutions correspond respectively to the minimum of the free
energy (\ref{eq:freeen}) and to the maximum of the entropy
(\ref{eq:entrop}). 

We now present results of the phase diagrams for the two ensembles in Fig.~\ref{fig:twoens}.
Both diagrams are qualitatively similar, with three phases:
paramagnetic, ferromagnetic, and nematic. For large values of the
parameter $\Delta$, on decreasing the energy/temperature, one observes a second-order transition from the
paramagnetic to the ferromagnetic phase; only at lower values of
$\Delta$ this phase transition becomes of first order. For low values
of $\Delta$, decreasing the energy/temperature results in a second-order
transition from the paramagnetic to the purely nematic phase for which $R_1$
is zero; a further decrease results in either a second-order
transition (for very small values of $\Delta$), or, 
a first-order transition (for $\Delta \approx 1/2$), to the ferromagnetic phase that has non-zero $R_1$.

\begin{figure}[h!]
\centering
\includegraphics[width=0.95\columnwidth]{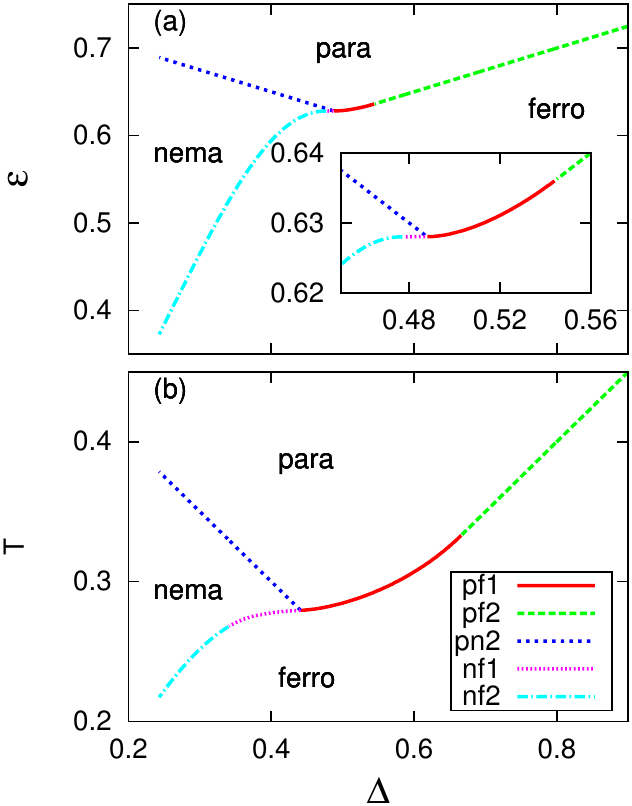}
\caption{(Color online) Comparison of the canonical and the microcanonical phase
diagram. Here, pf1 means first-order para-ferro transition, etc. 
(a) Phase diagram in the $\Delta$ -- $\e$ plane in the
microcanonical ensemble, Eqs.~(\ref{eq:scmfmc}) and (\ref{eq:entrop}). The two
tricritical points are at $\Delta\approx 0.545$, $\e\approx 0.636$, and at
$\Delta\approx 0.477$, $\e\approx 0.628$, while there is a critical end
point at $\Delta \approx 0.487$, $\e \approx 0.628$. The inset shows a zoom into
the central part.
 (b) Phase diagram in the $\Delta$ -- $T$ plane in the
canonical ensemble, Eqs.~(\ref{eq:scmfcan}) and (\ref{eq:freeen}). There
are two tricritical points at $\Delta\approx 0.667$, $T\approx 0.333$, and at $\Delta\approx 0.34$, 
$T\approx0.267$. The critical end point is at $\Delta \approx 0.441$,
$T \approx 0.279$.
}
\l{fig:twoens}
\end{figure}

While the phase diagrams in Fig.~\ref{fig:twoens} look simple, their mappings
onto each other (Fig.~\ref{fig0}) reveal nontrivial 
inequivalence between the canonical and microcanonical descriptions.
This inequivalence is because while the 
self-consistent solutions (\ref{eq:scmfcan},\ref{eq:cantd}) 
and (\ref{eq:scmfmc},\ref{eq:mcde}) are the same for both the ensembles
and transform onto one another by using
Eq.~(\ref{eq:beta-e}), they are nevertheless stable in different parameter regimes. Thus, using the mapping, Eq.~(\ref{eq:beta-e}), two
situations can arise: either a gap, i.e., a region of inaccessible
states, or an overlap, i.e., a region of multiple stable solutions.
Note that the second-order transition to the nematic phase is the same
in both the descriptions. 

\begin{figure}[th!]
\centering
\includegraphics[width=0.95\columnwidth]{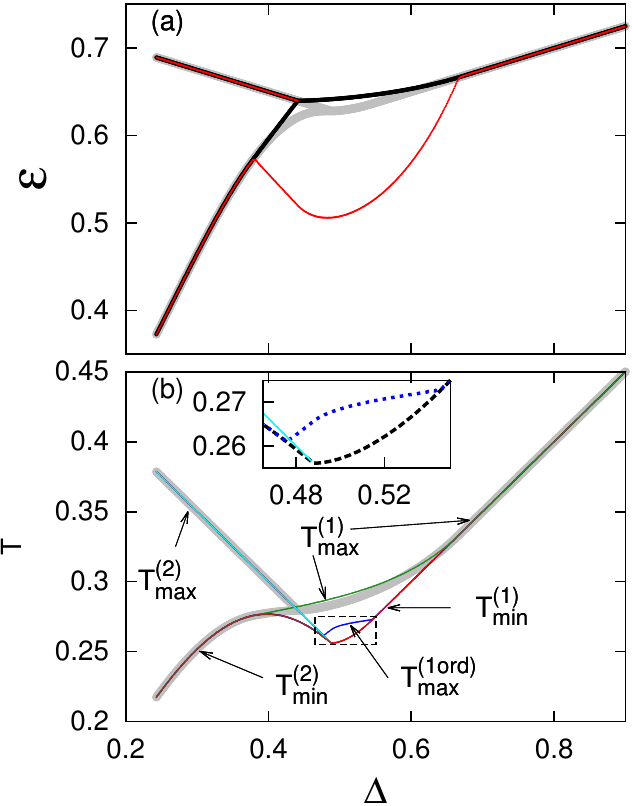}
\caption{(Color online)
Inequivalence of phase diagrams in the two ensembles.
  (a) Canonical phase diagram Fig.~\ref{fig:twoens}(b) mapped
onto the  $\Delta$ -- $\e$ plane (the microcanonical 
diagram is in background in gray). Between the bold black and the thin red lines, there is 
no canonical equilibrium state possible.
(b) Microcanonical phase diagram  Fig.~\ref{fig:twoens}(a) 
mapped onto the $\Delta$ -- $T$ plane 
(the canonical diagram is in the background in gray). 
$T^{(1)}_{\rm min}$ (red) is the minimal temperature 
at which the  paramagnetic phase exists. 
$T^{(1)}_{\rm max}$ (green) is the  maximum temperature at which the
ferromagnetic phase exists.
$T^{(2)}_{\rm max}$ (cyan) (respectively, $T^{(2)}_{\rm min}$ (brown)) 
is the maximum
(respectively, minimum) temperature at which the nematic phase exists.
The blue line for $T^{\rm (1ord)}_{\rm max}$ 
shows the splitting of the first-order microcanonical transition in the
region $0.477<\Delta<0.545$ (another line that belongs to this splitting
is masked by $T^{(1)}_{\rm min}$ and $T^{(2)}_{\rm min}$).
The inset shows a zoom into this middle region, where black dashed and blue
dotted lines correspond to the two values of the temperature
at the microcanonical jump.}
\l{fig0}
\end{figure}

As Fig.~\ref{fig0}(a) shows, mapping of the
canonical phase diagram onto the $\Delta$ -- $\e$ plane
yields a gap. In the domain of $\Delta$ where a first-order canonical transition occurs, the canonical transition line splits into two lines when mapped onto the $\Delta$ -- $\e$ plane.
Between these lines, there is no stable canonical state for a given $\e$ (cf. Fig.~\ref{fig2}).

A more nontrivial situation arises due to the mapping of the microcanonical phase diagram 
onto the $\Delta$ -- $T$ plane, as shown in Fig.~\ref{fig0}(b). Here,
two features are evident. First, in regions where the microcanonical transition 
is of second order but the canonical transition is of first order, there
are three microcanonically stable values of $R_{1,2}$ at temperatures between the lines $T^{(1)}_{\rm max}$ (green line) and
$T^{(1)}_{\rm min}$ (red line), and those between the lines
$T^{(1)}_{\rm max}$ and
$T^{(2)}_{\rm min}$ (brown line). Second, in regions of a first-order
microcanonical transition, the transition line splits into two lines,
denoted $T^{({\rm 1ord})}_{\rm max}$ (blue line) and $T^{({\rm
1ord})}_{\rm min}$ (black dashed line), with the latter coinciding with
either $T^{(1)}_{\rm min}$ or $T^{(2)}_{\rm min}$, such that
for temperatures in between, there are two microcanonically stable
values of $R_{1,2}$, see the inset of Fig.~\ref{fig0}(b) and cuts of the
$\Delta$ -- $T$ phase diagram at fixed values of $\Delta$ in Fig.~\ref{fig2}. Thus, in the whole domain
of $\Delta$ where the canonical transition is of first-order, one observes a multiplicity
of microcanonically stable states in the $\Delta$ -- $T$ plane. Remarkably, the tricritical points are different in the two ensembles. 

\begin{figure}
\centering
\includegraphics[width=0.65\columnwidth]{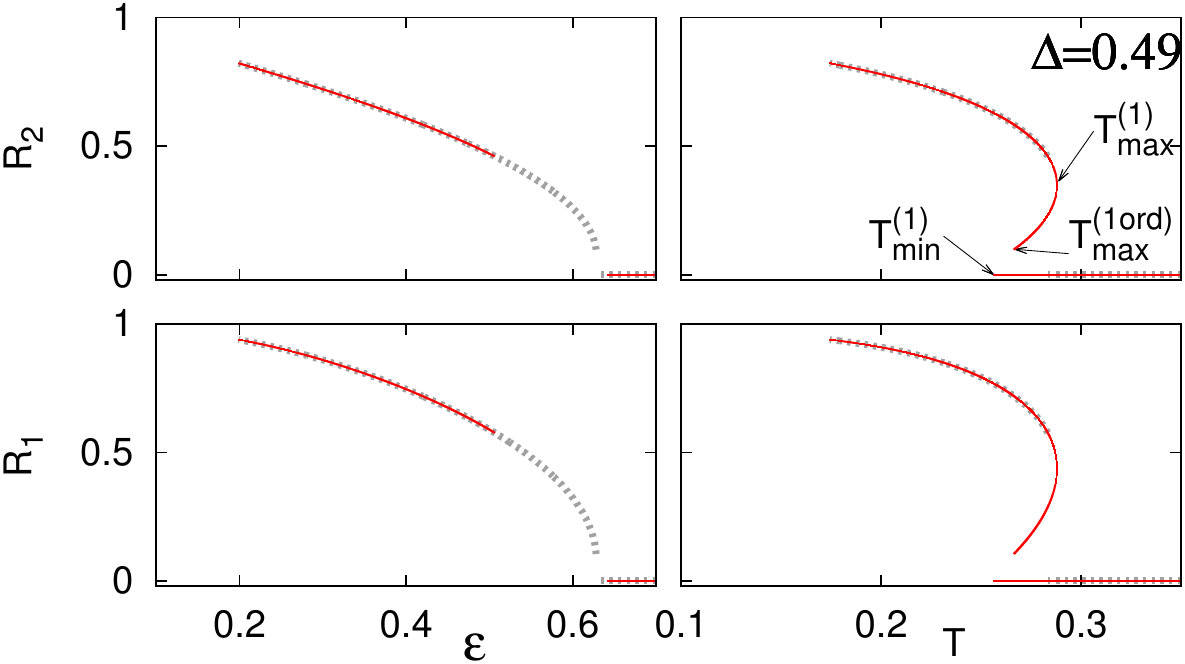}
\includegraphics[width=0.65\columnwidth]{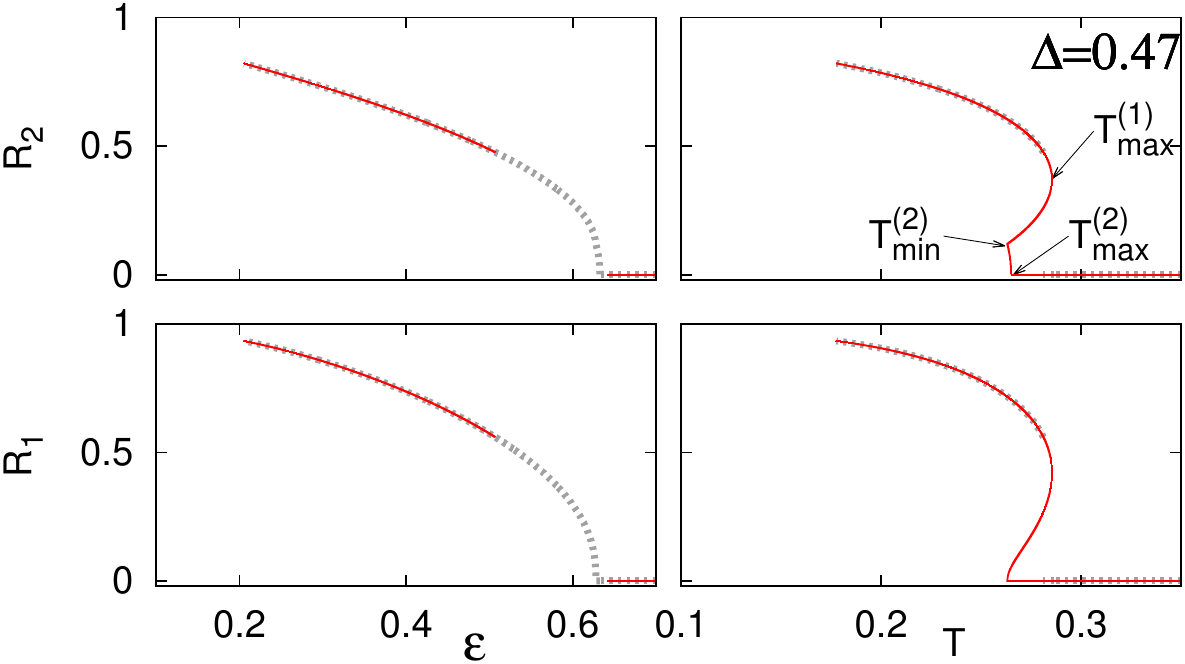}
\caption{(Color online) Stable solutions of $R_{1,2}$ vs. temperature
$T$ in the canonical ensemble and energy $\epsilon$ in the
microcanonical ensemble (dotted grey lines); red solid lines are stable
``imports'' from another ensemble (canonically stable states on left
column panels and microcanonically stable states on right column
panels); $\Delta$ equals $0.49$ (top panel), and $0.47$ (bottom panel). The values
of $T^{(1)}_{\rm max}, T^{(1)}_{\rm min}, T^{(2)}_{\rm max},
T^{(2)}_{\rm min}, T^{({\rm 1ord})}_{\rm max}$ marked by arrows
coincide with those in Fig.~\ref{fig0}.}
\l{fig2}
\end{figure}

In Fig. \ref{fig4}, we employ relation (\ref{eq:beta-e}) to draw the
temperature-energy relation $T(\e)$ for $\Delta=0.5$. Both for the
microcanonical and the canonical ensemble, this curve has two branches:
a high-energy branch, and a low-energy branch. At the point where the two branches intersect, the
two entropies in the microcanonical ensemble and the two free energies
in the canonical ensemble become equal. In the region where the
canonical curve shows a jump in the energy at a given temperature, characteristic of
a first-order transition that here occurs between the paramagnetic and
the ferromagnetic phase (see Fig. \ref{fig:twoens}(b)), the microcanonical
curve shows a region of negative specific heat ($\partial T/\partial \e
< 0$). Since the canonical specific heat is always positive, being given
by the fluctuations in the energy of the system, the negative
microcanonical specific heat is a further indication of ensemble
inequivalence for the model under study.

\begin{figure}
\centering
\includegraphics[width=0.45\columnwidth]{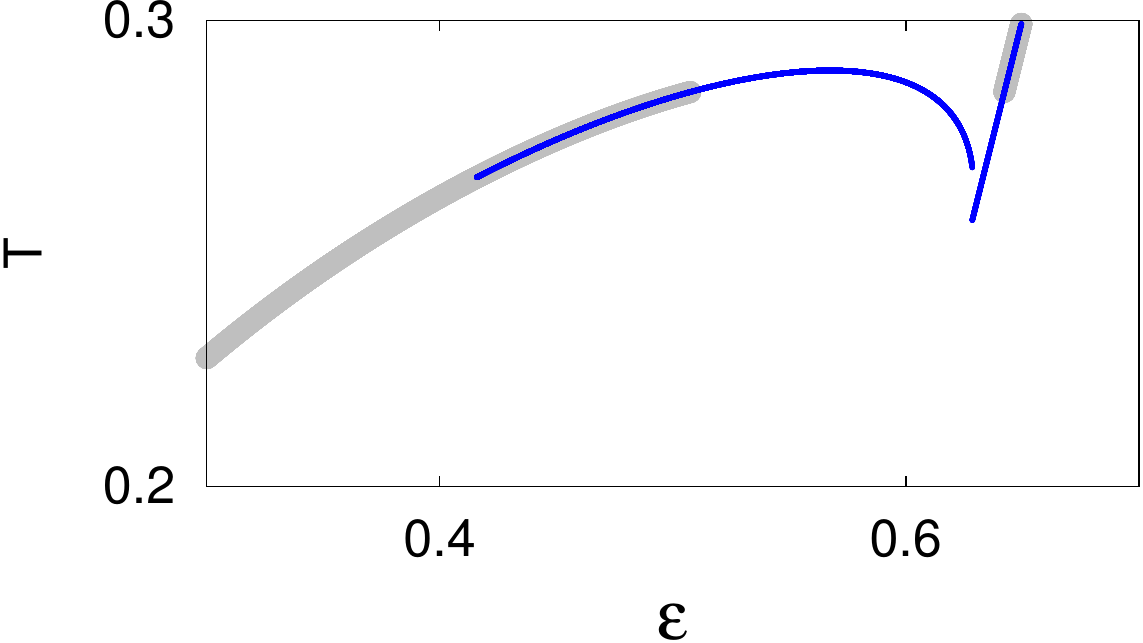}
\caption{(Color online) Plot of the dependence $\e$-$T$ for $\Delta=0.5$, showing
regions of microcanonical energies that are inaccessible canonically. Bold grey lines: canonically stable states, blue solid lines: microcanonically stable states.}
\l{fig4}
\end{figure}

To conclude, we addressed the issue of ensemble inequivalence
in long-range interacting systems, by studying an XY model of classical spins with linear and quadratic
coupling, and evolving under Hamilton dynamics. In
this so-called Generalized
Hamiltonian mean-field model, we compared exact equilibrium phase diagrams
in the microcanonical and canonical ensembles. We showed that within the region of
first-order transitions, the two ensembles show very different behaviors. Nevertheless, let us remark that when plotted using appropriate variables, the arrangement of critical points and transition
lines is similar in the phase diagrams of the two ensembles. One may study how the relaxation
to equilibrium differs in the two ensembles, a behavior investigated earlier in the microcanonical ensemble in Ref. \cite{Teles:2012}.
In that paper, it was shown that an isolated  system described by the
Hamiltonian (\ref{eq:H}) relaxes to quasi-stationary states (QSSs) which also
have paramagnetic, ferromagnetic, and nematic phases. The phase diagram of QSS, however, is very different from the one predicted by the equilibrium
statistical mechanics in the microcanonical ensemble, Fig.
\ref{fig:twoens}. Nevertheless, we expect that since the life-time of QSS scales with the number of particles in the system, a finite system will eventually relax to the Boltzmann-Gibbs equilibrium.  In the thermodynamic limit, however, this relaxation might take longer than the age of the Universe.
It will be of interest to explore such dynamical behavior in the canonical ensemble.

Finally, we mention that an overdamped nonequilibrium version of the GHMF
is a Kuramoto-type model of synchronization of globally coupled
oscillators (just as an overdamped nonequilibrium version of the HMF model is the standard Kuramoto
model~\cite{Gupta:20141,Gupta:20142}), where transitions to synchronization are of major interest. 
In the context of synchronization, nematic and ferromagnetic
phases correspond respectively to two-cluster and one-cluster
synchronization patterns (see Ref.~\cite{Komarov:2013}), 
but their stability is obtained from dynamical and not from free
energy/entropy considerations.

AP is supported by the grant from the agreement of August 27,
2013, number 02.В.49.21.0003 between the Ministry of Education and Science of
the Russian Federation and the Lobachevsky State University of Nizhni Novgorod.
SG is supported by the CEFIPRA Project 4604-3. The authors thank the Galileo Galilei Institute for Theoretical
Physics, Florence, Italy for the hospitality and the INFN for partial support during the
completion of this work. SG and AP acknowledge useful discussions with M.
Komarov. This work was partially supported by the CNPq, FAPERGS,
INCT-FCx, and by the US-AFOSR under the grant FA9550-12-1-0438.


\begin{thebibliography}{99}
\bibitem{Campa:2009}A. Campa, T. Dauxois and S. Ruffo, Phys. Rep. {\bf
480}, 57 (2009).

\bibitem{Bouchet:2010}F. Bouchet, S. Gupta and D. Mukamel, Physica A
{\bf 389}, 4389 (2010).

\bibitem{Levin:2014}Y. Levin, R. Pakter, F. B. Rizzato, T. N. Teles and
F. P. C. Benetti, Phys. Rep. {\bf 535}, 1 (2014).

\bibitem{Campa:2014}A. Campa, T. Dauxois, D. Fanelli and S. Ruffo, {\it 
Physics of Long-Range Interacting Systems} (Oxford University Press,
Oxford, 2014).

\bibitem{Sopik:2005}J. Sopik, C. Sire and P. H. Chavanis, Phys. Rev. E {\bf 72}, 026105 (2005).

\bibitem{Nicholson:1992}D. R. Nicholson, {\it Introduction to Plasma
Physics} (Krieger Publishing Company, Florida, 1992).

\bibitem{Landau:1960}L. D. Landau and E. M. Lifshitz {\it
Electrodynamics of Continuous Media} (Pergamon, London, 1960).

\bibitem{Chavanis:2002}P. H. Chavanis {\it Dynamics and Thermodynamics
of Systems with Long-range Interactions (Lecture Notes in Physics vol.
602)} edited by T. Dauxois, S. Ruffo, E. Arimondo and M. Wilkens (Springer-Verlag,
Berlin, 2002).

\bibitem{Paddy:1990}T. Padmanabhan, Phys. Rep. {\bf 188}, 285 (1990).

\bibitem{Lynden-Bell:1968}D. Lynden-Bell and R. Wood, Mon. Not. R.
Astron. Soc. {\bf 138}, 495 (1968).

\bibitem{Thirring:1970}W. Thirring, Z. Phys. {\bf 235}, 339 (1970).

\bibitem{Kiessling:1997}M. K. H. Kiessling and J. L. Lebowitz, Lett.
Math. Phys. {\bf 42}, 43 (1997).

\bibitem{Barre:2001}J. Barr\'{e}, D. Mukamel and S. Ruffo, Phys. Rev.
Lett. {\bf 87}, 030601 (2001).

\bibitem{Ispolatov:2001}I. Ispolatov and E. G. D. Cohen, Physica A {\bf
295}, 475 (2001).

\bibitem{Barre:2007}J. Barr\'{e} and B. Gon\c{c}alves, Physica A {\bf 386}, 212 (2007).

\bibitem{Mukamel:2005}D. Mukamel, S. Ruffo and N. Schreiber, Phys. Rev.
Lett. {\bf 95}, 240604 (2005).

\bibitem{Venaille:2009}A. Venaille and F. Bouchet, Phys. Rev. Lett.
{\bf 102}, 104501 (2009).

\bibitem{Venaille:2011}A. Venaille and F. Bouchet, J. Stat. Phys. {\bf
143}, 346 (2011).

\bibitem{Teles:2014}T. N. Teles, D. Fanelli and Stefano Ruffo, Phys.
Rev. E {\bf 89}, 050101(R) (2014).


\bibitem{Bouchet:2005}F. Bouchet and J. Barr\'{e}, J. Stat. Phys. {\bf
118} 1073 (2005).

\bibitem{Bouchet:2008}F. Bouchet, T. Dauxois, D. Mukamel and S. Ruffo,
Phys. Rev. E {\bf 77}, 011125 (2008).

\bibitem{Ruffo:1995}M. Antoni and S. Ruffo, Phys. Rev. E {\bf 52}, 2361 (1995).

\bibitem{Yamaguchi:2004}Y. Y. Yamaguchi, J. Barr\'{e}, F. Bouchet, T.
Dauxois and S. Ruffo, Physica A {\bf 337}, 36 (2004).

\bibitem{Campa:2007}A. Campa, A. Giansanti and G. Morelli, Phys. Rev. E
{\bf 76}, 041117 (2007).

\bibitem{Joyce:2010}M. Joyce and T. Worrakitpoonpon, J. Stat. Mech.:
Theory Exp. P10012 (2010).

\bibitem{Lee:1985}D. H. Lee and G. Grinstein, Phys. Rev. Lett. {\bf 55}, 541 (1985).

\bibitem{Teles:2012}T. N. Teles, F. P. C Benetti, R. Pakter R and Y.
Levin, Phys. Rev. Lett. {\bf 109}, 230601 (2012).

\bibitem{Kac:1963}M. Kac, G. E. Uhlenbeck and P. C. Hemmer, J. Math.
Phys. {\bf 4}, 216 (1963).

\bibitem{Qi:2013} K. Qi, M. H. Qin, X. T. Jia and J.-M. Liu, J. Magn. Magn. Mater.,
{\bf 340}, 127 (2013).

\bibitem{Carpenter:1989} D. B. Carpenter and J. T. Chalker, J. Phys.: Condens. Matter,
{\bf 1}, 4907 (1989)

\bibitem{Park:2008} J.-H. Park, S. Onoda, N. Nagaosa and J. H. Han, Phys. Rev. Lett.,
{\bf 101}, 167202 (2008)

\bibitem{Antoni:2002}M. Antoni, S. Ruffo and A. Torcini, Phys. Rev. E {\bf 66}, 025103(R) (2002).

\bibitem{Komarov:2013} M. Komarov and A. Pikovsky, Phys. Rev. Lett. {\bf 111}, 204101 (2013);
Physica D {\bf 289}, 18 (2014); V. Vlasov, M. Komarov and A. Pikovsky,
arXiv:1411.3204.

\bibitem{Gupta:20141}S. Gupta, A. Campa and S. Ruffo, Phys. Rev. E {\bf
89}, 022123 (2014).

\bibitem{Gupta:20142}S. Gupta, A. Campa and S. Ruffo, J. Stat. Mech.:
Theory Exp. R08001 (2014).
\end{thebibliography}
\end{document}